\newif\ifpreprint
\preprintfalse % Enable for double column reprint

\ifpreprint
\documentclass[journal=jpclcd,manuscript=letter]{achemso}
\else
\documentclass[journal=jpclcd,manuscript=letter,layout=twocolumn]{achemso}
\fi

\usepackage[T1]{fontenc} % Use modern font encodings

\usepackage{amsmath}
\usepackage{newtxtext,newtxmath}

\usepackage{pifont}
\usepackage{graphicx}
\usepackage{dcolumn}
\usepackage{braket}
\usepackage{multirow}
\usepackage{threeparttable}
\usepackage{xspace}
\usepackage{verbatim}
\usepackage[version=4]{mhchem} % Formula subscripts using \ce{}
\usepackage{comment}
\usepackage{color,soul}

\usepackage{mathtools}
\usepackage[dvipsnames]{xcolor}
\usepackage{xspace}
\usepackage{ifthen}

\usepackage{qcircuit}

\usepackage{graphicx,longtable,dcolumn,mhchem}
\usepackage{rotating,color}
\usepackage{lscape}
\usepackage{amsmath}
\usepackage{dsfont}
\usepackage{soul}
\usepackage{physics}

\newcommand{\cmark}{\color{green}{\text{\ding{51}}}}
\newcommand{\xmark}{\color{red}{\text{\ding{55}}}}

\newcommand{\mr}{\multirow}
\newcommand{\ie}{\textit{i.e.}}

\newcommand{\ra}{\rightarrow}

\newcommand{\SI}{\textcolor{blue}{supporting information}}

\newcommand{\SetA}{QUEST\#1}
\newcommand{\SetB}{QUEST\#2}
\newcommand{\SetC}{QUEST\#3}

\usepackage[colorlinks = true,
            linkcolor = blue,
            urlcolor  = black,
            citecolor = blue,
            anchorcolor = black]{hyperref}
\definecolor{goodorange}{RGB}{225,125,0}
\definecolor{goodgreen}{RGB}{5,130,5}
\definecolor{goodred}{RGB}{220,50,25}
\definecolor{goodblue}{RGB}{30,144,255}

\newcommand{\note}[2]{
\ifthenelse{\equal{#1}{F}}{
\colorbox{goodorange}{\textcolor{white}{\footnotesize \fontfamily{phv}\selectfont #1}}
    \textcolor{goodorange}{{\footnotesize \fontfamily{phv}\selectfont #2}}\xspace
}{}
\ifthenelse{\equal{#1}{R}}{
\colorbox{goodred}{\textcolor{white}{\footnotesize \fontfamily{phv}\selectfont #1}}
    \textcolor{goodred}{{\footnotesize \fontfamily{phv}\selectfont #2}}\xspace
}{}
\ifthenelse{\equal{#1}{N}}{
\colorbox{goodgreen}{\textcolor{white}{\footnotesize \fontfamily{phv}\selectfont #1}}
    \textcolor{goodgreen}{{\footnotesize \fontfamily{phv}\selectfont #2}}\xspace
}{}
\ifthenelse{\equal{#1}{M}}{
\colorbox{goodblue}{\textcolor{white}{\footnotesize \fontfamily{phv}\selectfont #1}}
    \textcolor{goodblue}{{\footnotesize \fontfamily{phv}\selectfont #2}}\xspace
}{}
}

\usepackage{titlesec}

\usepackage[fontsize=11pt]{scrextend}
\titleformat{\section}
{\normalfont\sffamily\bfseries\color{Blue}}
{\thesection.}{0.25em}{\uppercase}

\titleformat{\subsection}[runin]
{\normalfont\sffamily\bfseries}
{\thesubsection}{0.25em}{}[.\;\;]

\titleformat{\suppinfo}
{\normalfont\sffamily\bfseries}
{\thesubsection}{0.25em}{}

\titlespacing*{\section}{0pt}{0.5\baselineskip}{0.01\baselineskip}
\titlespacing*{\subsection}{0pt}{0.125\baselineskip}{0.01\baselineskip}

\author{Pierre-Fran\c{c}ois Loos}
	\email{loos@irsamc.ups-tlse.fr}
	\affiliation[LCPQ, Toulouse]{Laboratoire de Chimie et Physique Quantiques, Universit\'e de Toulouse, CNRS, UPS, France}
\author{Anthony Scemama}
	\affiliation[LCPQ, Toulouse]{Laboratoire de Chimie et Physique Quantiques, Universit\'e de Toulouse, CNRS, UPS, France}
\author{Denis Jacquemin}
	\email{Denis.Jacquemin@univ-nantes.fr}
	\affiliation[CEISAM, Nantes]{Universit\'e de Nantes, CNRS,  CEISAM UMR 6230, F-44000 Nantes, France}

\let\oldmaketitle\maketitle
\let\maketitle\relax
	\title{The Quest For Highly Accurate Excitation Energies: A Computational Perspective}
\date{\today}

\begin{tocentry}
	\vspace{1cm}
	\includegraphics[width=\textwidth]{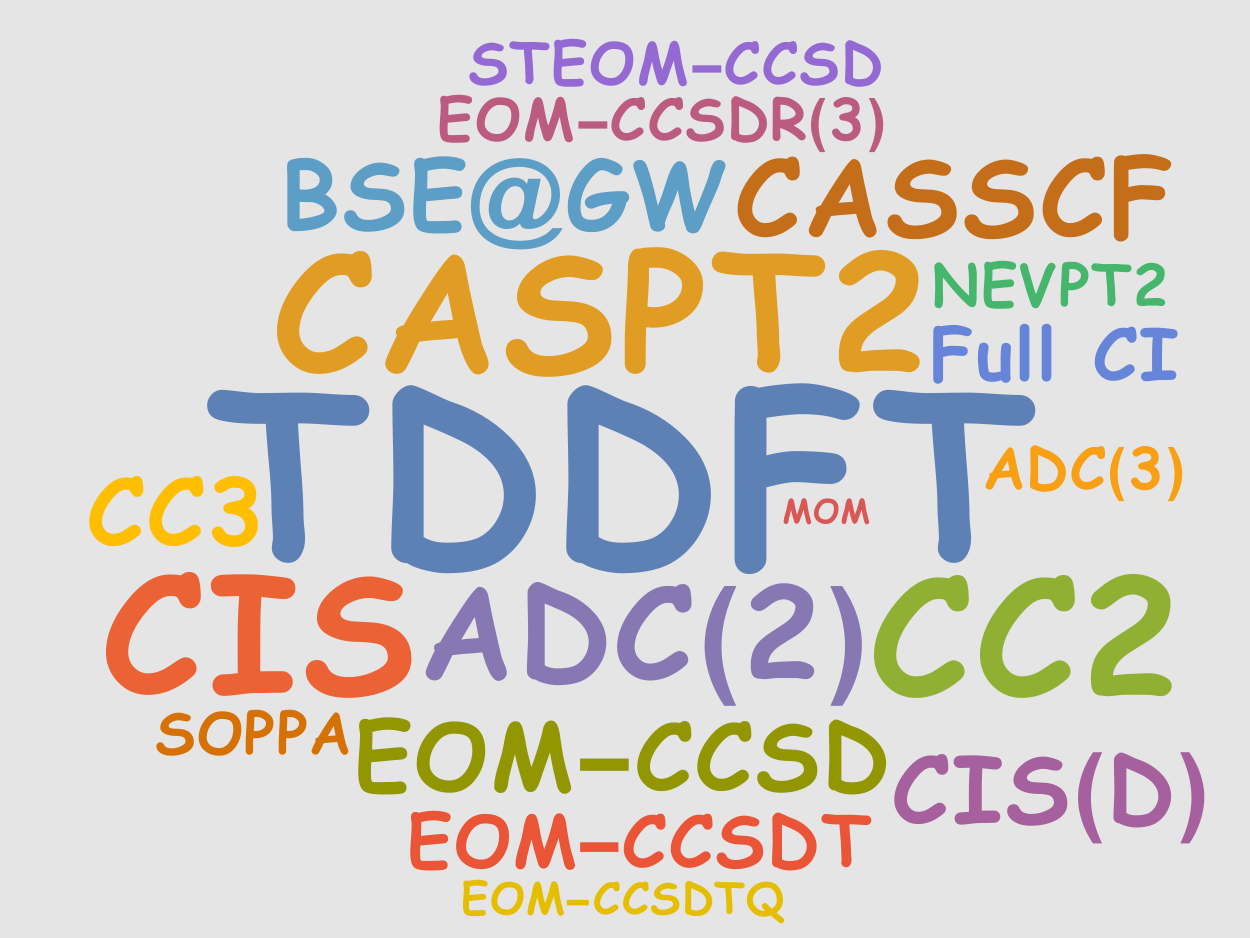}
\end{tocentry}

\begin{document}	

\ifpreprint
\else
\twocolumn[
\begin{@twocolumnfalse}
\fi
\oldmaketitle

%%%%%%%%%%%%%%%%
%%% ABSTRACT %%% 150 WORd MAX !!!!!!! 149 lˆ.
%%%%%%%%%%%%%%%%

\begin{abstract}
We provide an overview of the successive steps that made possible to obtain increasingly accurate excitation energies with computational chemistry tools, eventually leading to chemically accurate vertical transition energies for small- and medium-size molecules.     
First, we describe the evolution of \textit{ab initio} methods employed to define benchmark values, with originally Roos' CASPT2 method, then the CC3 method as in the renowned Thiel set, and more recently the resurgence of selected configuration interaction methods. 
The latter method has been able to deliver consistently, for both single and double excitations, highly accurate excitation energies for small molecules, as well as medium-size molecules with compact basis sets.
Second, we describe how these high-level methods and the creation of representative benchmark sets of excitation energies have allowed to assess fairly and accurately the performance of computationally lighter methods. 
We conclude by discussing the future theoretical and technological developments in the field.
\end{abstract}

\ifpreprint
\else
\end{@twocolumnfalse}
]
\fi

\ifpreprint
\else
\small
\fi

\noindent

%%%%%%%%%%%%%%%%%%%%
%%% INTRODUCTION %%%
%%%%%%%%%%%%%%%%%%%%
The accurate modeling of excited-state properties with \textit{ab initio} quantum chemistry methods is a clear ambition of the electronic structure theory community that will certainly keep us busy for (at the very least) the next few decades to come 
(see, for example, Refs.~\citenum{Dre05,Gon12,Gho18} and references therein). Of particular interest is the access to precise excitation energies, \ie, the energy difference between ground and excited electronic states, and their intimate link with photophysical and 
photochemical processes. The factors that makes this quest for high accuracy particularly delicate are very diverse.

First of all (and maybe surprisingly), it is, in most cases, tricky to obtain reliable and accurate experimental data that one can straightforwardly compare to theoretical values. In the case of vertical excitation energies, \ie, excitation energies at a fixed geometry, band maxima 
do not usually correspond to theoretical values as one needs to take into account both geometric relaxation and zero-point vibrational energy motion. Even more problematic, experimental spectra might not be available in gas phase, and, in the worst-case scenario, no clear 
assignment could be made. For a more faithful comparison between theory and experiment, although more computationally demanding, the so-called 0-0 energies are definitely a safer playground. \cite{Die04b,Win13,Fan14b,Loo19b}

Second, developing theories suited for excited states is usually more complex and costly than their ground-state equivalent, as one might lack a proper variational principle for excited-state energies. 
As a consequence, for a given level of theory, excited-state methods are usually less accurate than their ground-state counterpart, potentially creating a ground-state bias that leads to inaccurate excitation energies.

Another feature that makes excited states particularly fascinating and challenging is that they can be both very close in energy from each other and have very different natures ($\pi \ra \pi^*$, $n \ra \pi^*$, charge transfer, double excitation, valence, Rydberg, singlet, 
triplet, etc). Therefore, it would be highly desirable to possess a computational method (or protocol) that provides a balanced treatment of the entire ``spectrum'' of excited states. 
We think that, at this stage, none of the existing methods does provide such a feat at an affordable cost for chemically-meaningful compounds.

What are the requirement of the ``perfect'' theoretical model? As mentioned above, a balanced treatment of excited states with different character is highly desirable. Moreover, chemically accurate excitation energies (\ie, with error smaller than $1$~kcal/mol or $0.043$~eV) 
would be also beneficial in order to provide a quantitative chemical picture. The access to other properties, such as oscillator strengths, dipole moments, and analytical energy gradients, is also an asset if one wants to compare with experimental data.
Let us not forget about the requirements of minimal user input and minimal chemical intuition (\ie, black box models are preferable) in order to minimize the potential bias brought by the user appreciation of the problem complexity. Finally, low computational scaling with 
respect to system size and small memory footprint cannot be disregarded. Although the simultaneous fulfillment of all these requirements seems elusive, it is useful to keep these criteria in mind. Table \ref{tab:method} is here for fulfill such a purpose. 
In this Table, we also provide the typical error bar associated with each of these methods. 
Table S1 of the {\SI} reports additional details about (some of the) existing BSE and wave function theory benchmarks, whereas a review of TD-DFT benchmark studies
can be found elsewhere. \cite{Lau13} 
As can be seen in Table S1, the actual error bar obtained for a given method strongly depends on the actual type of excited states and compounds.
Hence, the values listed in Table \ref{tab:method} should be viewed as ``typical'' errors for organic molecules, nothing more.

%%% TABLE I %%%
\begin{table}
\footnotesize
\caption{Formal computational scaling of various excited-state methods with respect to the number of one-electron basis functions $N$ and the accessibility of various key properties in popular computational software packages.
For organic derivatives, the typical error range for single excitations is also provided as a qualitative indicator of the method accuracy.}
\label{tab:method}
\begin{tabular}{p{2.1cm}cccc}
	\hline
	\mr{2}{*}{Method}	&	Formal		&	Oscillator 	&	Analytical 		&	Typical 		\\
						&	scaling		&	strength	&	gradients		&	error (eV)		\\
	\hline
	TD-DFT				&	$N^4$		&	\cmark		&	\cmark			&	$0.2$--$0.4$$^a$	\\
	BSE@\textit{GW} 		&	$N^4$		&	\cmark		&	\xmark			&	$0.1$--$0.3$$^b$	\\
	\\
	CIS					&	$N^5$		&	\cmark		&	\cmark			&	$\sim 1.0$		\\
	CIS(D)				&	$N^5$		&	\xmark		&	\cmark			&	$0.2$--$0.3$	\\
	ADC(2)				&	$N^5$		&	\cmark		&	\cmark			&	$0.1$--$0.2$	\\
	CC2					&	$N^5$		&	\cmark		&	\cmark			&	$0.1$--$0.2$	\\
	\\
	ADC(3)				&	$N^6$		&	\cmark		&	\xmark			&	$0.2$	\\
	EOM-CCSD			&	$N^6$		&	\cmark		&	\cmark			&	$0.1$--$0.3$		\\
	\\
	CC3					&	$N^7$		&	\cmark		&	\xmark			&	$\sim 0.04$		\\
	\\
	EOM-CCSDT			&	$N^8$		&	\xmark		&	\xmark			&	$\sim 0.03$		\\
	EOM-CCSDTQ			&	$N^{10}$		&	\xmark		&	\xmark			&	$\sim 0.01$		\\
	\\
	CASPT2/NEVPT2		&	$N!$			&	\cmark		&	\cmark			&	$0.1$--$0.2$	\\
	SCI					&	$N!$			&	\xmark		&	\xmark			&	$\sim 0.03$		\\
	FCI					&	$N!$			&	\cmark		&	\cmark			&	$0.0$			\\
	\hline
\end{tabular}	
\begin{flushleft}
$^a${The error range is strongly functional and state dependent. The values reported here are for well-behaved cases;}
$^b${Typical error bar for singlet transitions. Larger errors are often observed for triplet excitations.}
\end{flushleft}
\end{table}

%**************
%** HISTORY **%
%**************
Before detailing some key past and present contributions aiming at obtaining highly accurate excitation energies, we start by giving a historical overview of the various excited-state \textit{ab initio} methods that have emerged in the last fifty years.
Interestingly, for pretty much every single method, the theory was derived much earlier than their actual implementation in electronic structure software packages and the same applies to the analytical gradients when  available.

%%%%%%%%%%%%%%%%%%%%%
%%% POPLE'S GROUP %%%
%%%%%%%%%%%%%%%%%%%%%
The first mainstream \textit{ab initio} method for excited states was probably CIS (configuration interaction with singles) which has been around since the 1970's. \cite{Ben71} 
CIS lacks electron correlation and therefore grossly overestimates excitation energies and wrongly orders excited states.
It is not unusual to have errors of the order of $1$ eV which precludes the usage of CIS as a quantitative quantum chemistry method.
Twenty years later, CIS(D) which adds a second-order perturbative correction to CIS was developed and implemented thanks to the efforts of Head-Gordon and coworkers. \cite{Hea94,Ish95} 
This second-order correction greatly reduces the magnitude of the error compared to CIS, with a typical error range of $0.2$--$0.3$ eV.
 
%%%%%%%%%%%%%%%%%%%
%%% ROOS' GROUP %%%
%%%%%%%%%%%%%%%%%%%
In the early 1990's, the complete-active-space self-consistent field (CASSCF) method \cite{Roo80,And90} and its second-order perturbation-corrected variant CASPT2 \cite{And92} (originally developed in Roos' group) became very popular.
This was a real breakthrough.
Although it took more than ten years to obtain analytical gradients, \cite{Cel03} CASPT2 was probably the first method that could provide quantitative results for molecular excited states of genuine photochemical interest. \cite{Roo96}
Nonetheless, it is of common knowledge that CASPT2 has the clear tendency of underestimating vertical excitation energies in organic molecules.
Driven by Angeli and Malrieu, \cite{Ang01} the creation of the second-order $n$-electron valence state perturbation theory (NEVPT2) method several years later was able to cure some of the main theoretical deficiencies of CASPT2.
For example, NEVPT2 is known to be intruder state free and size consistent.
The limited applicability of these multiconfigurational methods is mainly due to the need of carefully defining an active space based on the desired transition(s) in order to obtain meaningful results, as well as their factorial computational growth with the number of active electrons and orbitals.
With a typical minimal valence active space tailored for the desired transitions, the usual error with CASPT2 or NEVPT2 calculations is $0.1$--$0.2$ eV, with the additional complication of the possible IPEA correction for the former method. \cite{Zob17}
We also point out that some emergent approaches, like DMRG (density matrix renormalization group), \cite{Bai19} offer a new path for the development of these multiconfigurational methods.

%%%%%%%%%%%%%
%%% TDDFT %%%
%%%%%%%%%%%%%
The advent of time-dependent density-functional theory (TD-DFT) \cite{Run84,Cas95} was a significant step for the community as TD-DFT was able to provide accurate excitation energies at a much lower cost than its predecessors in a black-box way.
For low-lying valence excited states, TD-DFT calculations relying on hybrid exchange-correlation functionals have a typical error of $0.2$--$0.4$ eV.
However, a large number of shortcomings were quickly discovered. \cite{Toz98,Toz99,Dre04,Mai04,Dre05,Lev06,Eli11}
In the present context, one of the most annoying feature of TD-DFT --- in its most standard (adiabatic) approximation --- is its inability to describe, even qualitatively, charge-transfer states, \cite{Toz99,Dre04} Rydberg states, \cite{Toz98} and double excitations. \cite{Mai04,Lev06,Eli11}
These issues, as well as other well-documented shortcomings of DFT and TD-DFT, are related to the so-called delocalization error. \cite{Aut14a}
One closely related issue is the selection of the exchange-correlation functional from an ever-growing zoo of functionals and the variation of the excitation energies that one can observe with different functionals. \cite{Goe19,Sue19} 
More specifically, despite the development of new, more robust approaches (including the so-called range-separated \cite{Sav96,IIk01,Yan04,Vyd06} and double \cite{Goe10a,Bre16,Sch17} hybrids), it is still difficult (not to say impossible) to select a functional adequate for all families of transitions. \cite{Lau13} 
Moreover, the difficulty of making TD-DFT systematically improvable obviously hampers its applicability.
Despite all of this, TD-DFT remains nowadays the most employed excited-state method in the electronic structure community (and beyond).
 
%%%%%%%%%%%%%%%%%%
%%% CC METHODS %%%
%%%%%%%%%%%%%%%%%%
Thanks to the development of coupled cluster (CC) response theory, \cite{Koc90} and the growth of computational resources, equation-of-motion coupled cluster with singles and doubles (EOM-CCSD) \cite{Sta93} became mainstream in the 2000's.
EOM-CCSD gradients were also quickly available. \cite{Sta95}
With EOM-CCSD, it is not unusual to have errors as small as $0.1$ eV for small compounds and generally $0.2$ eV for larger ones, with a typical overestimation of the vertical transition energies.
Its third-order version, EOM-CCSDT, was also implemented and provides, at a significantly higher cost, high accuracy for single excitations. \cite{Kuc01}
Although extremely expensive and tedious to implement, higher orders are also technically possible for small systems thanks to automatically generated code. \cite{Kuc91,Hir04}
For the sake of brevity, we drop the EOM acronym in the rest of this \textit{Perspective} keeping in mind that these CC methods are applied to excited states in the present context.

The original CC family of methods was quickly completed by an approximated and computationally lighter family with, in front line, the second-order CC2 model \cite{Chr95} and its third-order extension, CC3. \cite{Chr95b}
As a $N^7$ method (where $N$ is the number of basis functions), CC3 has a particularly interesting accuracy/cost ratio with errors usually below the chemical accuracy threshold. \cite{Hat05c,Loo18a,Loo18b,Loo19a}
The series CC2, CCSD, CC3, CCSDT defines a hierarchy of models with $N^5$, $N^6$, $N^7$ and $N^8$ scaling, respectively.
It is also noteworthy that CCSDT and CC3 are also able to detect the presence of double excitations, a feature that is absent from both CCSD and CC2. \cite{Loo19c} 

%%%%%%%%%%%%%%%%%%%
%%% ADC METHODS %%%
%%%%%%%%%%%%%%%%%%%
It is also important to mention the recent rejuvenation of the second- and third-order algebraic diagrammatic construction [ADC(2) \cite{Sch82} and ADC(3) \cite{Tro99,Har14}] methods that scale as $N^5$ and $N^6$, respectively.
These methods are related to the older second- and third-order polarization propagator approaches (SOPPA and TOPPA). \cite{Odd78,Pac96}
This renaissance was certainly initiated by the enormous amount of work invested by Dreuw's group in order to provide a fast and efficient implementation of these methods, \cite{Dre15} including the analytical gradients, \cite{Reh19} as well as other interesting variants. \cite{Dre15,Hod19a}
These Green's function one-electron propagator techniques indeed represent valuable alternatives thanks to their reduced cost compared to their CC equivalents.
In that regard, ADC(2) is particularly attractive with an error around $0.1$--$0.2$ eV.
However, we have recently observed that ADC(3) generally overcorrects the ADC(2) excitation energies and is significantly less accurate than CC3. \cite{Tro02,Loo18a,Loo20a,Loo20b}

%%%%%%%%%%%%%%
%%% BSE@GW %%%
%%%%%%%%%%%%%%
Finally, let us mention the many-body Green's function Bethe-Salpeter equation (BSE) formalism \cite{Str88} (which is usually performed on top of a \textit{GW} calculation). \cite{Hed65}
BSE has gained momentum in the past few years and is a serious candidate as a computationally inexpensive electronic structure theory method that can effectively model excited states with a typical error of $0.1$--$0.3$ eV, as well as some related properties. \cite{Jac17b,Bla18}
One of the main advantage of BSE compared to TD-DFT (with a similar computational cost) is that it allows a faithful description of charge-transfer states and, when performed on top of a (partially) self-consistently \textit{GW} calculation, BSE@\textit{GW} has been shown to be weakly dependent on its starting point (\ie, on the functional selected for the underlying DFT calculation). \cite{Jac16,Gui18}
However, due to the adiabatic (\ie, static) approximation, doubly excited states are completely absent from the BSE spectrum.

%%%%%%%%%%%%%%%%%%%
%%% SCI METHODS %%%
%%%%%%%%%%%%%%%%%%%
In the past five years, \cite{Gin13,Gin15} we have witnessed a resurgence of the so-called selected CI (SCI) methods \cite{Ben69,Whi69,Hur73} thanks to the development and implementation of new, fast, and efficient algorithms to select cleverly 
determinants in the full CI (FCI) space (see Refs.~\citenum{Gar18,Gar19} and references therein). 
SCI methods rely on the same principle as the usual CI approach, except that determinants are not chosen \textit{a priori} based on occupation or excitation criteria but selected among the entire set of determinants based on their estimated contribution to the FCI wave function or energy. 
Indeed, it has been noticed long ago that, even inside a predefined subspace of determinants, only a small number of them significantly contributes.
The main advantage of SCI methods is that no \textit{a priori} assumption is made on the type of electron correlation. 
Therefore, at the price of a brute force calculation, a SCI calculation is not, or at least less, biased by the user appreciation of the problem's complexity.
One of the strength of one of the implementation, based on the CIPSI (configuration interaction using a perturbative selection made iteratively) algorithm developed by Huron, Rancurel, and Malrieu \cite{Hur73} is its parallel efficiency which makes possible to run on thousands of CPU cores. \cite{Gar19}
Thanks to these tremendous features, SCI methods deliver near FCI quality excitation energies for both singly and doubly excited states, \cite{Hol17,Chi18,Loo18a,Loo19c} with an error of roughly $0.03$ eV, mostly originating from the extrapolation procedure. \cite{Gar19}
However, although the \textit{``exponential wall''} is pushed back, this type of method is only applicable to molecules with a small number of heavy atoms and/or relatively compact basis sets.

%%%%%%%%%%%%%%%%%
%%% COMPUTERS %%%
%%%%%%%%%%%%%%%%%
For someone who has never worked with SCI methods, it might be surprising to see that one is able to compute near-FCI excitation energies for molecules as big as benzene. \cite{Chi18,Loo19c,Loo20a}
This is mainly due to some specific choices in terms of implementation as explained below.
Indeed, to keep up with Moore's ``Law'' in the early 2000's, the processor designers had no other choice than to propose multi-core chips to avoid an explosion of the energy requirements.
Increasing the number of floating-point operations per second by doubling the number of CPU cores only requires to double the required energy, while doubling the frequency multiplies the required energy by a factor of $\sim$ 8.
This bifurcation in hardware design implied a \textit{change of paradigm} \cite{Sut05} in the implementation and design of computational algorithms. A large degree of parallelism is now required to benefit from a significant acceleration.
Fifteen years later, the community has made a significant effort to redesign the methods with parallel-friendly algorithms. \cite{Val10,Cle10,Gar17b,Pen16,Kri13,Sce13}
In particular, the change of paradigm to reach FCI accuracy with SCI methods came
from the use of determinant-driven algorithms which were considered for long as inefficient
with respect to integral-driven algorithms.
The first important element making these algorithms efficient is the introduction of new bit manipulation instructions (BMI) in the hardware that enable an extremely fast evaluation of Slater-Condon rules \cite{Sce13b} for the direct calculation of
the Hamiltonian matrix elements over arbitrary determinants.
Then massive parallelism can be harnessed to compute the second-order perturbative correction with semi-stochatic algorithms, \cite{Gar17b,Sha17} and perform the sparse matrix multiplications required in Davidson's algorithm to find the eigenvectors associated with the lowest eigenvalues.
Block-Davidson methods can require a large amount of memory, and the recent introduction of byte-addressable non-volatile memory as a new tier in the memory hierarchy \cite{Pen19} will enable SCI calculations on larger molecules.
The next generation of supercomputers is going to generalize the presence of accelerators (graphical processing units, GPUs), leading to a new software crisis.
Fortunately, some authors have already prepared this transition. \cite{Dep11,Kim18,Sny15,Ufi08,Kal17}

%%%%%%%%%%%%%%%
%%% SUMMARY %%% 
%%%%%%%%%%%%%%%
In summary, each method has its own strengths and weaknesses, and none of them is able to provide accurate, balanced, and reliable excitation energies for all classes of electronic excited states at an affordable cost.

%%% FIG 1 %%%
\begin{figure*}
	\includegraphics[width=\linewidth]{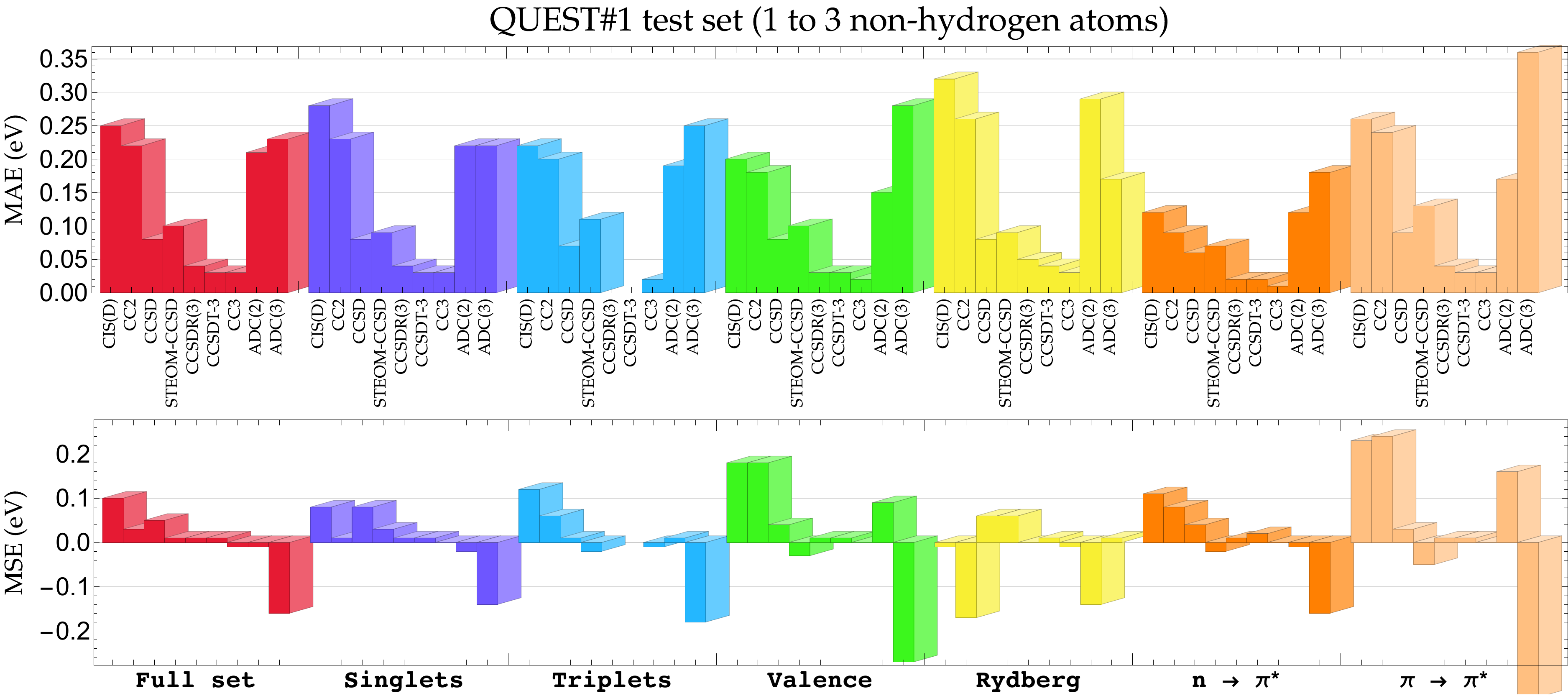}
	\caption{Mean absolute error (MAE, top) and mean signed error (MSE, bottom) with respect to the TBE/\textit{aug}-cc-pVTZ values from the {\SetA} set (as described in Ref.~\citenum{Loo18a}) for various methods and types of excited states. 
	The corresponding graph for the maximum positive and negative errors can be found in the {\SI}.
	}
	\label{fig:Set1}
\end{figure*}
%%% %%% %%%

%%% FIG 2 %%%
\begin{figure}
	\includegraphics[width=\linewidth]{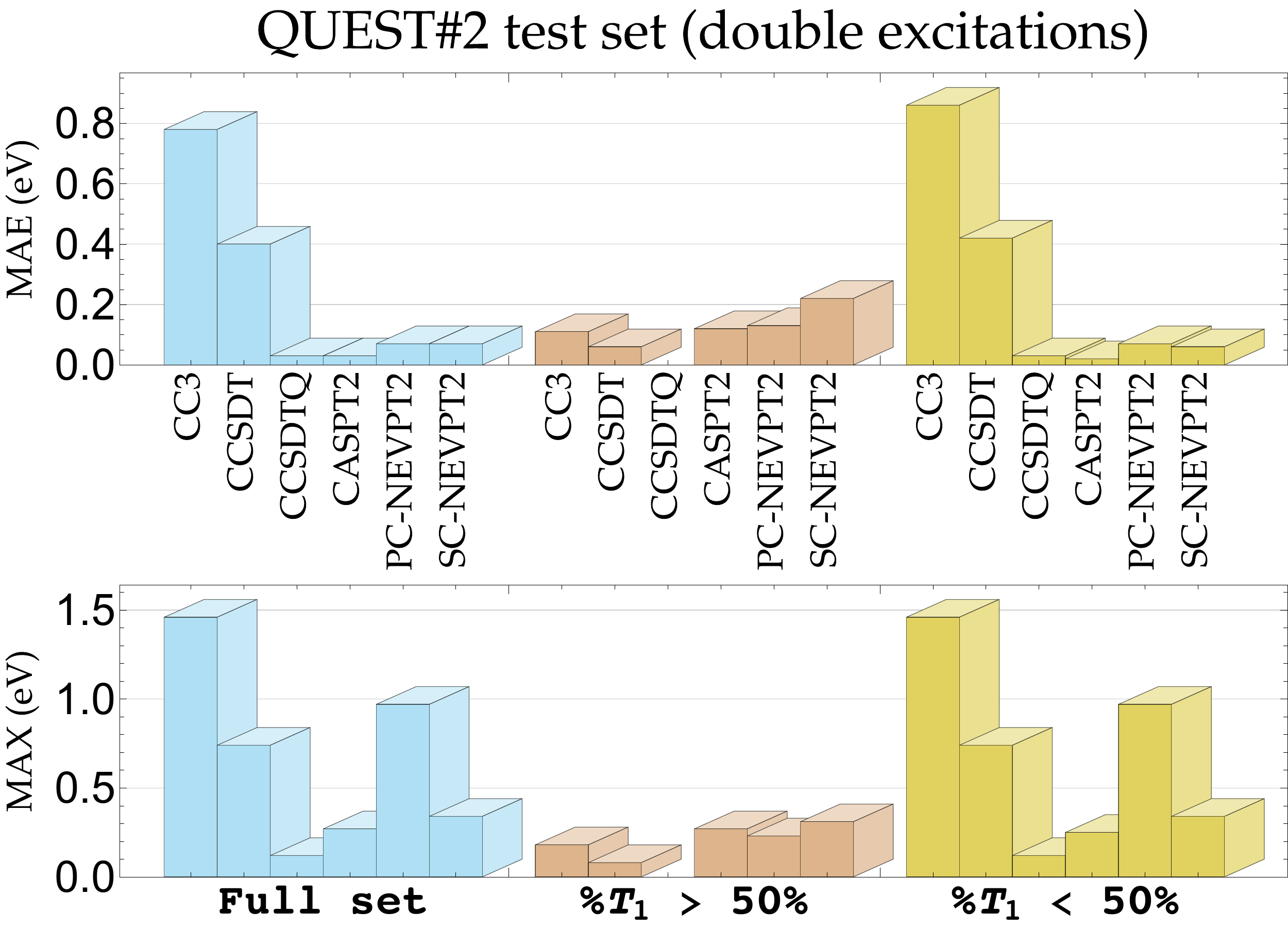}
	\caption{Mean absolute error (MAE, top) and maximum absolute error (MAX, bottom) with respect to FCI excitation energies for the doubly excited states reported in Ref.~\citenum{Loo19c} for various methods taking into account at least triple excitations.
	$\%T_1$ corresponds to single excitation percentage in the transition calculated at the CC3 level.
	For this particular set and methods, the mean signed error is equal to the MAE.}
	\label{fig:Set2}
\end{figure}
%%% %%% %%%

%%% FIG 3 %%%
\begin{figure*}
	\includegraphics[width=\linewidth]{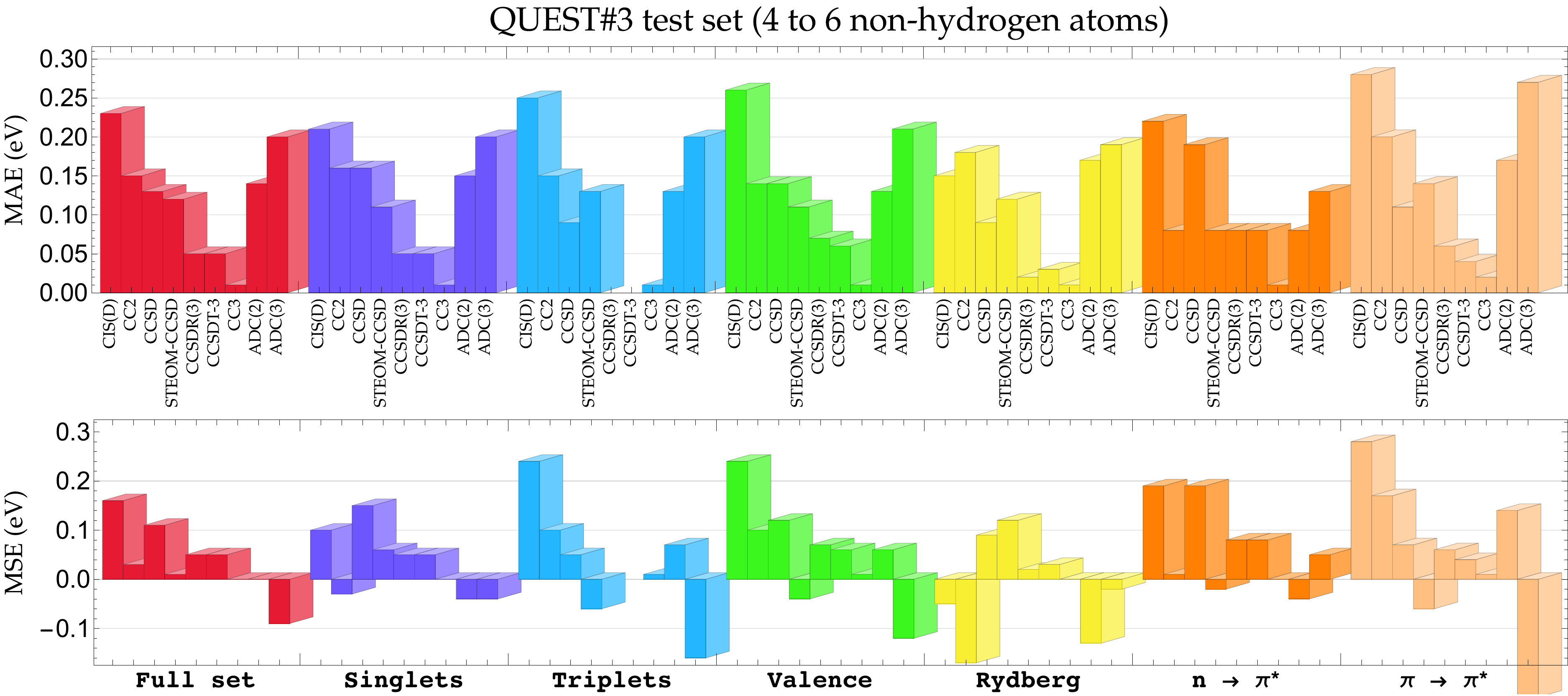}
	\caption{Mean absolute error (MAE, top) and mean signed error (MSE, bottom) with respect to the TBE/\textit{aug}-cc-pVTZ values from the {\SetC} set (as described in Ref.~\citenum{Loo20a}) for various methods and types of excited states.
	The corresponding graph for the maximum positive and negative errors can be found in the {\SI}.
	}
	\label{fig:Set3}
\end{figure*}
%%% %%% %%%

%*****************
%** BENCHMARKS ***
%*****************
Although sometimes decried, benchmark sets of molecules and their corresponding reference data are essential for the validation of existing theoretical models and to bring to light and subsequently understand their strengths and, more importantly, their limitations. 
These sets have started to emerge at the end of the 1990's for ground-state properties with the acclaimed G2 test set designed by the Pople group. \cite{Cur97}
For excited states, things started moving a little later but some major contributions were able to put things back on track.

%%%%%%%%%%%%%%%%%%%
%%% THIEL'S SET %%%
%%%%%%%%%%%%%%%%%%%
One of these major contributions was provided by the group of Walter Thiel \cite{Sch08,Sil08,Sau09,Sil10b,Sil10c} with the introduction of the so-called Thiel (or M\"ulheim) set of excitation energies. \cite{Sch08}
For the first time, this set was large, diverse, consistent, and accurate enough to be used as a proper benchmarking set for excited-state methods.
More specifically, it gathers a large number of excitation energies consisting of 28 medium-size organic molecules with a total of 223 valence excited states (152 singlet and 71 triplet states) for which theoretical best estimates (TBEs) were defined.
In their first study Thiel and collaborators performed CC2, CCSD, CC3 and CASPT2 calculations (with the TZVP basis) in order to provide (based on additional high-quality literature data) TBEs for these transitions.
Their main conclusion was that \textit{``CC3 and CASPT2 excitation energies are in excellent agreement for states which are dominated by single excitations''}.
These TBEs were quickly refined with the larger \textit{aug}-cc-pVTZ basis set, \cite{Sil10b} highlighting the importance of diffuse functions.
As a direct evidence of the actual value of reference data, these TBEs were quickly picked up to benchmark various computationally effective methods from semi-empirical to state-of-the-art \textit{ab initio} methods (see the \textit{Introduction} of Ref.~\citenum{Loo18a} and references therein).

Theoretical improvements of Thiel's set were slow but steady, highlighting further its quality. \cite{Wat13,Kan14,Har14,Kan17} 
In 2013, Watson \textit{et al.} \cite{Wat13} computed CCSDT-3/TZVP (an iterative approximation of the triples of CCSDT \cite{Wat96}) excitation energies for the Thiel set.
Their quality were very similar to the CC3 values reported in Ref.~\citenum{Sau09} and the authors could not appreciate which model was the most accurate.
Similarly, Dreuw and coworkers performed ADC(3) calculations on Thiel's set and arrived at the same kind of conclusion: \cite{Har14}
\textit{``based on the quality of the existing benchmark set it is practically not possible to judge whether ADC(3) or CC3 is more accurate''}.
These two studies clearly demonstrate and motivate the need for higher accuracy benchmark excited-states energies. 

%%%%%%%%%%%%%%%%%%%%%%%
%%% JACQUEMIN'S SET %%%
%%%%%%%%%%%%%%%%%%%%%%%
Recently, we made, what we think, is a significant contribution to this quest for highly accurate vertical excitation energies. \cite{Loo18a}
More specifically, we studied 18 small molecules with sizes ranging from one to three non-hydrogen atoms.
For such systems, using a combination of high-order CC methods, SCI calculations and large diffuse basis sets, we were able to compute a list of 110 highly accurate vertical excitation energies for excited states of various natures (valence, Rydberg, $n \ra \pi^*$, $\pi \ra \pi^*$, singlet, triplet and doubly excited) based on CC3/\textit{aug}-cc-pVTZ geometries.
In the following, we label this set of TBEs as {\SetA}.
Importantly, it allowed us to benchmark a series of popular excited-state wave function methods partially or fully accounting for double and triple excitations (see Fig.~\ref{fig:Set1}): CIS(D), CC2, CCSD, STEOM-CCSD, \cite{Noo97} CCSDR(3), \cite{Chr77} CCSDT-3, \cite{Wat96} CC3, ADC(2), and ADC(3).
Our main conclusion was that CC3 is extremely accurate (with a mean absolute error of only $\sim 0.03$ eV), and that, although slightly less accurate than CC3, CCSDT-3 could be used as a reliable reference for benchmark studies. 
Quite surprisingly, ADC(3) was found to have a clear tendency to overcorrect its second-order version ADC(2). 
The mean absolute errors (MAEs) obtained for this set can be found in Fig.~\ref{fig:Set1}.

In a second study, \cite{Loo19c} using a similar combination of theoretical models (but mostly extrapolated SCI energies), we provided accurate reference excitation energies for transitions involving a substantial amount of double excitations using a series of increasingly large diffuse-containing atomic basis sets (up to \textit{aug}-cc-pVQZ when technically feasible). 
This set gathers 20 vertical transitions from 14 small- and medium-sized molecules, a set we label as {\SetB} in the remaining of this \textit{Perspective}.
An important addition to this second study was the inclusion of various flavors of multiconfigurational methods (CASSCF, CASPT2, and NEVPT2) in addition to high-order CC methods including, at least, perturbative triples (see Fig.~\ref{fig:Set2}).
Our results clearly evidence that the error in CC methods is intimately related to the amount of double-excitation character in the vertical transition. 
For ``pure'' double excitations (\ie, for transitions which do not mix with single excitations), the error in CC3 and CCSDT can easily reach $1$ and $0.5$ eV, respectively, while it goes down to a few tenths of an electronvolt for more common transitions (such as in \textit{trans}-butadiene and benzene) involving a significant amount of singles.\cite{Shu17,Bar18b,Bar18a}
The quality of the excitation energies obtained with multiconfigurational methods was harder to predict as the overall accuracy of these methods is highly dependent on both the system and the selected active space.
Nevertheless, CASPT2 and NEVPT2 were found to be more accurate for transition with a small percentage of single excitations (error usually below $0.1$ eV) than for excitations dominated by single excitations where the error is closer from $0.1$--$0.2$ eV (see Fig.~\ref{fig:Set2}).

In our latest study, \cite{Loo20a} in order to provide more general conclusions, we generated highly accurate vertical transition energies for larger compounds with a set composed by 27 organic molecules encompassing from four to six non-hydrogen atoms for a total of 223 vertical transition energies of various natures.
This set, labeled as {\SetC} and still based on CC3/\textit{aug}-cc-pVTZ geometries, is constituted by a reasonably good balance of singlet, triplet, valence, and Rydberg states.
To obtain this new, larger set of TBEs, we employed CC methods up to the highest possible order (CC3, CCSDT, and CCSDTQ), very large SCI calculations (with up to hundred million determinants), as well as the most robust multiconfigurational method, NEVPT2. 
Each approach was applied in combination with diffuse-containing atomic basis sets.
For all the transitions of the {\SetC} set, we reported at least CCSDT/\textit{aug}-cc-pVTZ (sometimes with basis set extrapolation) and CC3/\textit{aug}-cc-pVQZ transition energies as well as CC3/\textit{aug}-cc-pVTZ oscillator strengths for each dipole-allowed transition. 
Pursuing our previous benchmarking efforts, \cite{Loo18a,Loo19c} we confirmed that CC3 almost systematically delivers transition energies in agreement with higher-level theoretical models ($\pm 0.04$ eV) except for transitions presenting a dominant double excitation character (see Fig.~\ref{fig:Set3}).
This settles down, at least for now, the debate by demonstrating the superiority of CC3 (in terms of accuracy) compared to methods like CCSDT-3 or ADC(3), see Fig.~\ref{fig:Set3}.
Moreover, thanks to the exhaustive and detailed comparisons made in Ref.~\citenum{Loo20a}, we could safely conclude that CC3 also regularly outperforms CASPT2 (which often underestimates excitation energies) and NEVPT2 (which typically overestimates excitation energies) as long as the corresponding transition does not show any strong multiple excitation character.

Our current efforts are now focussing on expanding and merging these sets to create an complete test set of highly accurate excitations energies.
In particular, we are currently generating reference excitations energies for radicals as well as more ``exotic'' molecules containing heavier atoms (such as \ce{Cl}, \ce{P}, and \ce{Si}).
The combination of these various sets would potentially create an ensemble of more than 400 vertical transition energies for small- and medium-size molecules based on accurate ground-state geometries.
Such a set would likely be a valuable asset for the electronic structure community.
It would likely stimulate further theoretical developments in excited-state methods and provide a fair ground for the assessments of the currently available and under development excited-state models.

%%%%%%%%%%%%%%%%%%
%%% Properties
%%%%%%%%%%%%%%%%%%
Besides all the studies described above aiming at reaching chemically accurate vertical transition energies, it should be pointed out that an increasing amount of effort is currently devoted to the obtention of highly-trustable excited-state properties. 
This includes, first, 0-0 energies, \cite{Die04b,Hat05c,Goe10a,Sen11b,Win13,Fan14b,Loo18b,Loo19a,Loo19b} which, as mentioned above, offer well-grounded comparisons with experiment. 
However, because 0-0 energies are fairly insensitive to the underlying molecular geometries, \cite{Sen11b,Win13,Loo19a} they are not a good indicator of their overall quality. 
Consequently, one can find in the literature several sets of excited-state geometries obtained at various levels of theory, \cite{Pag03,Gua13,Bou13,Tun16,Bud17} some of them being determined using state-of-the-art models. \cite{Gua13,Bud17} 
There are also investigations of the accuracy of the nuclear gradients at the Franck-Condon point. \cite{Taj18,Taj19} 
The interested reader may find useful several investigations reporting sets of reference oscillator strengths. \cite{Sil10c,Har14,Kan14,Loo18a,Loo20b} 
Up to now, these investigations focusing on geometries and oscillator strengths have been mostly based on theory-vs-theory comparisons. Indeed, while for small compounds (\ie, typically from di- to tetra-atomic molecules), 
one can find very accurate experimental measurements (excited state dipole moments, oscillator strengths, vibrational frequencies, etc), these data are usually not accessible for larger compounds. 
Nevertheless, the emergence of X-ray free electron lasers might soon allow to obtain accurate experimental excited state densities and geometrical structures through diffraction experiments. 
Such new experimental developments will likely offer new opportunities for experiment-vs-theory comparisons going beyond standard energetics.
Finally, more complex properties, such as two-photon cross-sections and vibrations, have been mostly determined at lower levels of theory, hinting at future studies on this particular subject.

%%%%%%%%%%%%%%%%%%
%%% CONCLUSION %%%
%%%%%%%%%%%%%%%%%%
As concluding remarks, we would like to highlight once again the major contribution brought by Roos' and Thiel's groups in an effort to define benchmark values for excited states. 
Following their footsteps, we have recently proposed a larger, even more accurate set of vertical transitions energies for various types of excited states (including double excitations). \cite{Loo18a,Loo19c,Loo20a}
This was made possible thanks to a technological renaissance of SCI methods which can now routinely produce near-FCI excitation energies for small- and medium-size organic molecules. \cite{Chi18,Gar18,Gar19}
We hope that new technological advances will enable us to push further, in years to come, our quest to highly accurate excitation energies, and, importantly, of other excited-state properties.

%%%%%%%%%%%%%%%%%%%%%%%%
%%% ACKNOWLEDGEMENTS %%%
%%%%%%%%%%%%%%%%%%%%%%%%
PFL would like to thank Peter Gill for useful discussions.
He also acknowledges funding from the \textit{``Centre National de la Recherche Scientifique''}.
DJ acknowledges the \textit{R\'egion des Pays de la Loire} for financial support. 

%%%%%%%%%%%%%%%%%%%%
%%% BIBLIOGRAPHY %%%
%%%%%%%%%%%%%%%%%%%%
\bibliography{ExPerspective}

\end{document}